\documentstyle[prb,aps,12pt,preprint]{revtex}

\begin{document}
\draft
\title{Doppler shift on local density of states and local
impurity scattering in the vortex state}
\author{E. Schachinger}
\address{Institut f\"ur Theoretische Physik, Technische Universit\"at
Graz\\A-8010 Graz, Austria}
\author{J.P. Carbotte}
\address{Department of Physics and Astronomy, McMaster University,\\
Hamilton, Ont. L8S 4M1, Canada}
\date{\today}
\maketitle
\begin{abstract}
The vortex state thermal and transport properties of the high
$T_c$ copper oxides can be understood in a $d$-wave gap model
and are dominated by the extended quasiparticle states that exist
along the nodal directions in momentum space. The Doppler
shift on these states due to the circulating
supercurrents around the vortex core, introduces new van Hove ridges
into the energy dependent local density of states (LDOS)
as a function of distance in the region between cores.
We emphasize the topology of these ridges and the effect on them of
local static impurity scattering in Born and unitary limit. We treat
possible orthorhombicity. Effective local scattering rates are also
obtained.
\end{abstract}
\newpage
\section{Introduction}

The theoretical predictions of Volovik\cite{Volo1,Volo2} and
Kopnin and Volovik\cite{Kopn1,Kopn2} that the low temperature
specific heat in a $d$-wave superconductor would vary as the square
root of the applied magnetic field $(\sqrt{H})$ was verified
in several experiments.\cite{Moler1,Moler2,Junod,Revaz,Wright}
To understand the experimental data it is assumed that the extended
quasiparticle states outside the vortex cores can be treated
semiclassically taking account only of the Doppler shift due to the
circulating supercurrents. To obtain the specific heat, which is
a bulk property of the system, it is necessary to carry out
a spacial average over the vortex unit cell of the local
density of states (LDOS) which varies as a function of distance
from the vortex core center. The low temperature limit of the
specific heat depends on the zero frequency limit of this
averaged density of quasiparticle states. This quantity is not only
affected by the magnetic field {\bf H} but can also depend on
impurity content,\cite{Kub1,Schach} particularly in the resonant
scattering limit. In this limit the theory of a $d$-wave
superconductor without magnetic field predicts a finite, constant
value for its zero frequency limit which is proportional to the
self consistent effective scattering rate and this leads to the
well known universal limit for the transport.\cite{Lee,Graf,Wu,Taill}

When the magnetic field is oriented in the copper oxide plane
rather than perpendicular to it new anisotropy effects in the
specific heat are predicted.\cite{Schach,Vek1,Vek2,Hirsch1}
In addition transport properties are also affected by the Doppler shift of
a vortex state.\cite{Hirsch1,Kub2} The thermal conductivity has
been particularly widely studied.\cite{Kub2,Palstra,Peacor,%
Krish,Franz,Aubin,Chiao}

The semiclassical Doppler shift of the electron due to the
circulating supercurrents outside the vortex core should also
affect the frequency dependence of the LDOS which can be
measured in scanning tunneling microscope (STM) experiments.
In view of the above successes in our understanding of the thermal and
transport properties of the copper oxides it is of interest to understand the
signature of the Doppler shift on the LDOS and the effect of
impurities on it. Very recently Franz and
Te\v{s}anovi\'c\cite{Franz1} considered the LDOS problem and
compared an average over vortex winding angles of the
semi classical LDOS with
equivalent results obtained from complete solutions of the BdG
equations for a single vortex. Their main conclusion was
that the semiclassical approach does indeed provide a good approximation
for this quantity.

For no magnetic field
$(H = 0)$ there is a single van Hove singularity in the density
of states at energy $\omega$ equal to the gap amplitude which does
not shift with position ${\bf r}$ (homogeneous case). 
For finite $H$ new van Hove ridges are
predicted. The actual number of such ridges and their topology
depends on the winding angel $\beta$ and simple formulas can be
obtained from which their contour in $(\omega,{\bf r})$-space
can be traced.
The effect of impurities on the topology of these ridges
is considered as is the effect of orthorhombicity. Impurities are
treated in both Born and unitary limit, and the influence of the
circulating supercurrents on the local effective impurity scattering
is also considered. This topic was recently discussed by
Barash and Svidzinskii.\cite{Barash} Orthorhombicity, present in
YBa$_2$Cu$_3$O$_{6.95}$ (YBCO) because of the existence of chains, 
is treated within a simple
effective mass model for the in-plane electronic dispersion curves
and a possible subdominant $s$-wave component is added to the dominant
$d$-wave gap.

The tetragonal case is the subject of section II. Impurities are included.
Generalization to the orthorhombic case is given in section III.
Section IV deals with effective local impurity scattering rates. A
short conclusion is found in section V.

\section{Tetragonal case}

In a semiclassical approximation the local Green's function at
position {\bf r} is given in terms of the Doppler shift
${\bf v}_F({\bf k}){\bf q}_s$ by
\begin{equation}
 G({\bf k},\omega_n;{\bf r}) =
 -{\left[i\omega_n-{\bf v}_F({\bf k}){\bf q}_s\right]\tau_0 +
  \varepsilon_{\bf k}\tau_3+\Delta_{\bf k}\tau_1\over
  \left[\omega_n+i{\bf v}_F({\bf k}){\bf q}_s\right]^2+
  \varepsilon_{\bf k}^2 +\Delta^2_{\bf k}}
 \label{eq:1}
\end{equation}
where $\tau_{1,2,3}$ are the $2\times 2$ Pauli matrices, $i\omega_n$
the Matsubara frequencies $i(2n+1)\pi T, n = 0,\pm 1,\pm 2,\ldots$
and $T$ is the temperature. The electronic dispersion in momentum space
{\bf k} is $\varepsilon_{\bf k}$ and the gap in the two-dimensional
copper oxide Brillouin zone is $\Delta_{\bf k}$. The electronic
Fermi velocity is ${\bf v}_F({\bf k})$ and ${\bf q}_s$ is the
momentum associated with the superfluid flow about the vortex. In the
magnetic field region $H_{c1} < \vert{\bf H}\vert \ll H_{c2}$ with 
$H_{c1,2}$ the lower and upper critical field respectively, and {\bf H} the
external magnetic field taken to be perpendicular
to the CuO$_2$-planes, the intervortex distance $R = {1\over a}
\sqrt{\phi_0\over\pi \vert{\bf H}\vert}$ where $a$ is a geometrical
factor of order one which is associated with the vortex arrangement,
and $\phi_0$ is the fundamental quantum of flux. In what follows, the position
{\bf r} will be measured relative to the center of the vortex core.
We use polar coordinates $(r,\beta)$ with $\beta$ the vortex winding
angle. The dimensionless variable $\rho = r/R$ takes on the value
1 at the boundary of the vortex unit cell. It is expected that
$R \gg \xi_0$ ($\xi_0$ is the coherence length) and we will be
interested only in values of $r > \xi_0$.

Assuming a circular velocity field for the supercurrents around a
single vortex, ${\bf v}_s = {\hbar\hat{\beta}\over 2mr}$ with
$\hat{\beta}$ a unit vector along the current, the Doppler shift
\begin{equation}
 {\bf v}_F({\bf k}){\bf q}_s = {E_H\over\rho}\sin(\phi-\beta),
 \label{eq:2}
\end{equation}
where $E_H$ is the magnetic energy scale that enters our problem and
is equal to ${a\over 2}v_F\sqrt{\pi H\over \phi_0} \equiv \nu\Delta_0$
where the last identity measures $E_H$ in units of the gap
amplitude $\Delta_0$. In Eq.~(\ref{eq:2}) $\phi$ is the polar
angle associated with momentum {\bf k} on the Fermi surface and
the $d$-wave gap $\Delta_{\bf k} = \Delta_0\cos 2\phi \equiv
\Delta_d(\phi)$ in the same model.

The local quasiparticle density of states at relative position
$\rho$ and energy $\omega$, $N_L(\rho,\omega)$, follows
directly from the analytic continuation of $G$ defined in Eq.~(\ref{eq:1})
to real frequencies $i\omega_n\to\omega+i0^+$ with
\begin{equation}
{N_L(\rho,\omega)\over N_0} = \int\limits_0^{2\pi}\!{d\phi\over 2\pi}\,
 \Re{\rm e}\left\{{\left\vert\omega-{\bf v}_F({\bf k}){\bf q}_s\right\vert\over
  \sqrt{\left[\omega - {\bf v}_F({\bf k}){\bf q}_s\right]^2-
  \Delta_d^2(\phi)}}\right\},
\label{eq:3}
\end{equation}
where $N_0$ is the normal state density of states. Note that the
Doppler shift of Eq.~(\ref{eq:2}) which enters (\ref{eq:3}) depends
on the magnetic field $\vert{\bf H}\vert = H$ through
$E_H \sim \sqrt{H}$ and on the position of the STM tip away
from the vortex core given by $\rho$ and $\beta$. We will
be interested only in the region where {\bf r} is outside
the vortex core of size $\xi_0$. This is mandated by the
fact that our approach  does not treat in detail the core
interior but instead includes only the supercurrents outside the
core through the superfluid velocity field
${\bf v}_s = {\hbar\hat{\beta}\over 2mr}$ which itself is an
approximation. Nevertheless, it is the aim of this work
to understand how the presence of ${\bf v}_s$ modifies the
LDOS in the region between the vortex cores.

The integral in Eq.~(\ref{eq:3}) is easily evaluated. We
present a first set of results which include some impurity
scattering in Born approximation. This smooths out the
logarithmic singularity at the gap amplitude predicted
for pure $d$-wave with $H = 0$. To include static
impurity scattering it is simply necessary to exchange the
frequency $\omega$ and the gap $\Delta_d(\phi)$ in Eq.~(\ref{eq:3})
by their renormalized values denoted by a tilde, i.e.:
$\tilde{\omega}$ and $\tilde{\Delta}(\phi)$. The
renormalized frequency
\begin{equation}
  \tilde{\omega}(\rho,\beta,\omega) = \omega+
  i\pi t^+\Omega(\rho,\beta,\omega)
  \label{eq:4a}
\end{equation}
in Born approximation with $t^+$ the impurity scattering rate
in the normal state, and
\begin{equation}
  \tilde{\omega}(\rho,\beta,\omega) =
  \omega+i\pi\Gamma^+{\Omega(\rho,\beta,\omega)\over
   \Omega^2(\rho,\beta,\omega)+D^2(\rho,\beta,\omega)}
  \label{eq:4b}
\end{equation}
in the unitary limit with $\Gamma^+$ replacing $t^+$ of
Eq.~(\ref{eq:4a}). The renormalized gap
$\tilde{\Delta}(\phi) = \Delta_0\cos2\phi+i\tilde{\Delta}_s(\rho,\beta,\omega)$
in a $(d+is)$-wave symmetry with
\begin{equation}
   \tilde{\Delta}_s(\rho,\beta,\omega) = i\pi t^+D(\rho,\beta,\omega)
   \label{eq:5a}
\end{equation}
for Born scattering and
\begin{equation}
   \tilde{\Delta}_s(\rho,\beta,\omega) = i\pi\Gamma^+{D(\rho,\beta,\omega)
   \over\Omega^2(\rho,\beta,\omega)+D^2(\rho,\beta,\omega)}
   \label{eq:5b}
\end{equation}
in the unitary limit. Here, the two functions $\Omega(\rho,\beta,\omega)$
and $D(\rho,\beta,\omega)$ are
\begin{equation}
\Omega(\rho,\beta,\omega) = 
  \left\langle
  {\tilde{\omega}(\rho,\beta,\omega)-{\bf v}_F({\bf p}){\bf q}_s\over
  \sqrt{\left[\tilde{\omega}(\rho,\beta,\omega)-{\bf v}_F({\bf p}){\bf q}_s
   \right]^2-\tilde{\Delta}^2_s(\rho,\beta,\omega)-\Delta^2_d(\phi)}}
  \right\rangle_\phi
  \label{eq:6a}
\end{equation}
and
\begin{equation}
D(\rho,\beta,\omega) =
  \left\langle
  {\tilde{\Delta}_s(\rho,\beta,\omega)\over
  \sqrt{\left[\tilde{\omega}(\rho,\beta,\omega)-{\bf v}_F({\bf p}){\bf q}_s
   \right]^2-\tilde{\Delta}^2_s(\rho,\beta,\omega)-\Delta^2_d(\phi)}}
  \right\rangle_\phi,
  \label{eq:6b}
\end{equation}
where $\langle\cdots\rangle_\phi$ indicates an angular average over the
angles $\phi$ around the circular Fermi surface in the two-dimensional
copper oxide Brillouin zone.

In the top frame of Fig.~\ref{f1} we show our results for a $d$-wave superconductor
(tetragonal symmetry) with a gap amplitude $\Delta_0 = 24\,$meV.
We take the magnetic field ${\bf H}\parallel c$-axis, i.e. {\bf H} is perpendicular to
the copper oxygen planes and the vortex winding angle $\beta = 0^\circ$.
The vertical axis is labeled by $N_L(\rho,\omega)$ which is given by
\begin{equation}
   N_L(\rho,\omega) = \Re{\rm e}\left\{\Omega(\rho,\beta,\omega)\right\}_
   {\beta = const},
\end{equation}
and features the
LDOS as a function of $\rho$ and $\omega$. Estimates of the magnetic
energy scale $E_H$ in YBCO have been considered by Vekhter et al.\cite{Vek2}
This quantity is not very well known but best estimates give between
\[
  E_H = 20-30 [{\rm K}]\sqrt{H [{\rm Tesla}]}.
\]
 Therefore the magnetic energy scale
has been set at 10\% of the gap amplitude for illustrative purposes,
i.e. $\nu = 0.1$ in our notation and the static impurity content in
Born approximation is $t^+ = 0.1\,$meV, enough to smear out the van Hove
singularity but not enough to make them disappear.
(This value of $t^+$ corresponds to a scattering rate of about
$6\,$K. Recent experiments by Hosseini et al.\cite{Hoss} seem to
indicate an even smaller scattering rate.) Three prominent sets of ridges are
seen for $\rho$ in the range 0.1 (at the vortex core, i.e. at $r = \xi_0$)
and 1.0 (at the vortex cell boundary). The case with no magnetic field is
shown as the bottom frame of Fig.~\ref{f1} for comparison. Only one
prominent van Hove ridge is seen in this second case and it is centered at $\omega = 
\Delta_0 = 24\,$meV
and, of course, there is no $\rho$ dependence since there is no vortex core
and thus no spatial inhomogeneities in the problem. A main ridge around
$\omega = 24\,$meV is still seen in the top frame although as $\rho$ is
decreased and the vortex core is approached at $\rho = 0.1$ it has
shifted in energy significantly. The other two ridges are due entirely to the
Doppler shift provided by the supercurrents and can be understood as follows.
In the pure case (no impurity scattering) the ridges in $N_L(\rho,\omega)$
correspond to extrema in the zeros of the denominator of Eq.~(\ref{eq:3}),
i.e. they correspond to the extrema of the equation
\begin{equation}
 \omega = {\bf v}_F({\bf k}){\bf q}_s\pm\Delta(\phi) =
  \Delta_0\left[{\nu\over\rho}\sin(\phi-\beta)\pm\cos2\phi\right],
 \label{eq:7}
\end{equation}
i.e. to solutions of equation
\begin{equation}
 {d\omega\over d\phi} = 0 = {\nu\over\rho}\cos(\phi-\beta)\mp 2\sin2\phi,
 \label{eq:7a}
\end{equation}
which have been written for a general value of vortex winding angle
$\beta$. There are three solutions of this equation for the case considered
in the top frame of Fig.~\ref{f1} $(\beta = 0)$. 
The critical values of frequency $\omega_{ci}$
with $i=1,2,3$ are shown as a function of distance from the vortex
core $\rho$ in the top frame of Fig.~\ref{f2}. For
$\omega\ge 0$ the solutions occur at the critical angles
$\phi_{c1,2} = \pm\pi/2$ and $\phi_{c3} = \sin^{-1}(\nu/4\rho)$ 
respectively and the corresponding values of $\omega_{ci}$ are
\begin{equation}
  \omega_{c1,2} = \Delta_0(1\pm\nu/\rho),
 \label{7b}
\end{equation}
which are valid without restriction; 
in the limit $\nu/\rho \to 0$ we find
\begin{equation}
  \omega_{c3} = \Delta_0\left[1+{1\over 2}\left(
{\nu\over 2\rho}\right)^2\right].
  \label{7c}
\end{equation}
Numerical solutions
can always be obtained and are needed except for a few simple
cases. For $\nu = 0$, no magnetic field,
all $\omega_{ci} = \Delta_0$ as expected. At the vortex cell boundary
$\rho = 1$, i.e. $r = R$, the three critical frequencies are in
units of $\Delta_0$ 1.1, 0.9, and 1.0013 respectively, and at the
vortex core which corresponds to $\rho = 0.1$ for $\nu = 0.1$ these
values are 2, 0, and 1.125. These expectations are bourn out in
the top frame of Fig.~\ref{f1}. We note from Eq.~(\ref{7b}) that the
spacing of the two additional structures away from $\omega = \Delta_0$
goes in the antinodal direction $(\beta = 0^\circ)$ like
\[
   \omega_{c1,2} \propto \sqrt{H},
\]
i.e. is linear in $\nu$; the shift of the main van Hove ridge from $\Delta_0$
due to the Doppler shift goes instead like
\[
   \omega_{c3} \propto H,
\]
i.e. is linear in the magnetic field in the limit $\nu/\rho\to 0$.
Other directions, $\beta\ne 0^\circ$, will develop other $H$-dependencies
of the spacing of the van Hove ridges which will have to be evaluated
numerically in most cases.

Also note the $\omega\to 0$ limit
of the LDOS. It goes to zero linearly in the $H = 0$ case but,
in contrast, it takes on a finite value when vortices are present.
Its value depends strongly on position $\rho$ in the vortex unit cell
and increases strongly with decreasing $\rho$. It is the average
of the LDOS at $\omega = 0$ over the vortex unit cell, i.e. the
average over $\rho$ and winding angle $\beta$ that determines
the low temperature bulk specific heat and this varies with the
value of H. Franz and Te\v{s}anovi\'c\cite{Franz1} discuss in detail
how this region of small $\omega$ can be strongly modified by an
in-plane angular
dependence to the tunneling matrix element that comes into the
relationship between STM measurements and the LDOS itself. These
are only the same when the tunneling element is constant.

In the bottom frame of Fig.~\ref{f2} we show again the same
data as shown in the top frame of Fig.~\ref{f1} but in a different
representation which
allows to see the ridges a little more clearly. What is presented
here is $N_L(\rho,\omega)$ vs. $\omega$ for fixed values of $\rho$,
namely $\rho = 0.2$ (solid line), $\rho = 0.3$ (dashed line),
$\rho = 0.6$ (dotted line), and $\rho = 1.0$ (dash-dotted line). The vortex
core radius equal to the coherence length $\xi_0$ corresponds to
$\rho = 0.1$ and the inter vortex distance is $\rho = 1.0$. Finally,
a thin solid line indicates $N_L(\omega)$ for no magnetic field,
$H = \nu = 0$.

Furthermore, the data shows clearly the main van Hove singularity around, but
not precisely at $\omega \simeq \Delta_0$, with some variation with
$\rho$ as detailed in the top frame of Fig.~\ref{f2} (solid line,
$\omega_{c3}$). This peak
exists even when $H=0$ (light solid line) when there are no supercurrents and
it is the usual $d$-wave gap peak. Note that here we have included some
small amount of impurity scattering in Born approximation, namely
$t^+ = 0.1\,$meV. The other secondary peaks above and below
$\omega\simeq 24\,$meV are due to the supercurrents and are the
semiclassical signature of the resulting Doppler shifts. 
(Similar ridges with somewhat different geometry appear in
earlier work which makes use of the semiclassical Eilenberger
approach.\cite{Scho,Ichi1,Ichi2,Wong} This approach should be
better inside the core, a region we do not treat properly.
Other related work in the $s$-wave case is the very recent paper by
Eschrig et al.\cite{Eschrig}) As the
STM tip is moved more towards the vortex core, the secondary peak 
at low energy becomes fairly prominent and also the $\omega = 0$
limit becomes finite as noted before. The topology of the surface
$N_L(\rho,\omega)$ vs. $\rho$ and $\omega$ is affected by several external
variables such as the magnitude of the magnetic field, the vortex
winding angle, the type of impurities involved, i.e. Born or
unitary scattering, and its strength as we now detail.

The ridge structure in the LDOS changes considerably as the
position of the STM tip is changed. This is illustrated in the
top frame of Fig.~\ref{f3} where we show our result
for a winding angle $\beta = 45^\circ$, i.e. in the nodal direction.
There are striking differences with the top frame of Fig.~\ref{f1}.
Only two ridges are now seen in contrast to three in the $\beta = 0^\circ$
case (the antinodal direction), and the main gap peak around $\omega = 24\,$meV
at the vortex cell boundary $\rho = 1$ (at $r = R$) moves downward in
energy as $\rho$ is reduced toward the vortex core. The $\omega = 0$
value of $N_L(\rho,\omega)$ near $\rho = 0.1$, i.e. $r = \xi_0$, is also
much smaller than in the $\beta = 0^\circ$ case.

In the bottom frame if Fig.~\ref{f3} we show our solutions for the
critical values of $\omega$ ($\omega_{ci}, i = 1,2$) which locate the
ridges in the top frame.
The critical angles $\phi_{ci}, i = 1,2$ at which the extrema
responsible for the ridges occur are also presented as 
functions of the distance from the vortex center $\rho$ in this frame. It is
obvious that geometry strongly affects the topology of the
$N_L(\rho,\omega)$ surface.

Impurity scattering can also affect things in an important way.
In Fig.~\ref{f4} we show results for the LDOS $N_L(\rho,\omega)$ as a
function of the distance to the vortex core $\rho$ and energy
$\omega$ for a magnetic energy $\nu = 0.1$, ${\bf H}\parallel c$-axis,
and a winding angle $\beta = 0^\circ$ (antinodal direction). This
figure is to be compared directly with the top frame of Fig.~\ref{f1}
as the only difference between the two figures is that in Fig.~\ref{f3}
the unitary impurity limit is used, Eqs.~(\ref{eq:4b}) and
(\ref{eq:5b}), with $\Gamma^+ = 0.1\,$meV instead of the Born limit,
Eqs.~(\ref{eq:4a}) and (\ref{eq:5a}), with $t^+ = 0.1\,$meV. The
structure of the ridges is now more pronounced and the zero frequency
limit of the LDOS near the vortex cell boundary is now larger.

\section{Orthorhombicity}

YBCO is orthorhombic because of the existence of chains along the
$b$-axis in one of the three copper oxide two-dimensional planes
per unit cell. This implies
that the gap is in principle a mixture of $s$- and $d$-wave symmetry
which do belong to the same irreducible representation of the
crystal lattice group, and therefore has the
form
\begin{equation}
   \Delta(\phi) = \Delta_0\left(\cos 2\phi+s\right)
   \label{eq:o1}
\end{equation}
on the Fermi surface. Here $s$ is the subdominant $s$-wave component.
A simple way to include band anisotropy is
to introduce anisotropic effective masses\cite{Schach,Schur} with
$m_b < m_a$ in an infinite band model. This gives an ellipsoidal
Fermi surface which can always be mapped onto a circular one by
scaling of momentum variables. At the same time the gap is transformed
to\cite{Schach,Schur}
\begin{equation}
 \Delta(\phi) = \Delta_0\left(s +{\alpha +\cos2\phi\over
   1+\alpha\cos2\phi}\right).
 \label{eq:o2}
\end{equation}
There are no additional changes and the calculations proceed as before
with Eqs.~(\ref{eq:7}) and (\ref{eq:7a}). For the direction
$\beta = 0^\circ$, they are replaced by:
\begin{equation}
 \omega = \Delta_0\left[{\nu\over\rho}{1+\alpha\over\sqrt{1-\alpha}}
  \sin\phi\pm\left({\alpha +\cos2\phi\over
   1+\alpha\cos2\phi} + s\right)\right],
 \label{eq:o3}
\end{equation}
and
\begin{equation}
 {d\omega\over d\phi} = 0 = {\nu\over\rho}{1+\alpha\over\sqrt{1-\alpha}}
 \cos\phi\mp{2(1-\alpha^2)\sin2\phi\over(1+\alpha\cos2\phi)^2},
 \label{eq:o4}
\end{equation}
with $\alpha$ the effective mass anisotropy parameter
defined as\cite{Schach,Schur} $\alpha = (m_b-m_a)/(m_a+m_b)$. Again,
for a general winding angle numerical solutions are needed although
analytic ones are still possible in the limit $\nu/\rho\to 0$ in some
cases.  If we restrict ourselves to
solutions of Eq.~(\ref{eq:o2}) which have at least some positive values
for the critical frequency $\omega_c$ in the range $\nu\le\rho\le 1$
then we get for $\beta = 0^\circ$, and
the critical angles $\phi_c = \pm\pi/2$ the following three solutions
of Eq.~(\ref{eq:o3})
\begin{equation}
  {\omega_c\over\Delta_0} = \left\{\begin{array}{ll}
  {\nu\over\rho}{1+\alpha\over
  \sqrt{1-\alpha}} \pm (1-s), & \phi_c = {\pi\over 2}\\
  -{\nu\over\rho}{1+\alpha\over
   \sqrt{1-\alpha}}+(1-s), & \phi_c = -{\pi\over 2}
  \end{array} \right .
  \label{eq:o5}
\end{equation}
without restrictions on the value of $\nu/\rho$, and another solution
\begin{equation}
  {\omega_c\over\Delta_0} = 1+s+{1\over 2}\left({\nu\over 2\rho}\right)^2
   {(1+\alpha)^3\over(1-\alpha)^2},
  \label{eq:o6}
\end{equation}
which is valid only for $\nu/\rho\to 0$ and corresponds in that case
to $\phi_c$ near zero radians.

First we note the well known limit of $\nu = 0$ (no magnetic field)
in which case the van Hove singularity in the density of states is
split by the existence of the subdominant $s$-wave component of the
gap and there are two singularities in the density of states
positioned at $\omega_c =
\Delta_0(1\pm s)$. This holds whatever the value of the effective
mass anisotropy $\alpha$ might be. This mass anisotropy does not lead
to a corresponding splitting of the van Hove singularities in the
density of states. In a previous paper Sch\"urrer et al.\cite{Schur}
have considered the specific case of optimally doped YBCO and have
suggested that reasonable model parameters are $\alpha = 0.4$ to
account for a factor of a little more than 2 in the effective
conductivity between $b$- and $a$-direction. Further consideration
of finite temperature penetration depth data\cite{Bonn} suggests
$s = -0.25$. This gives agreement between the model and the measured
penetration depth data at low but finite temperatures $T$.

In the top frame of Fig.~\ref{f5} we show results for $N_L(\rho,\omega)$
 for a case with
$\alpha = 0.4$, $s = -0.25$, and $\beta = 0^\circ$, i.e. in the antinodal
direction. The top frame is a three dimensional plot of $N_L(\rho,\omega)$
vs. $\rho$ and $\omega$ and is to be compared with the top frame
of Fig.~\ref{f1} (tetragonal case). We note striking differences in
the ridge structure. Now, the two ridges at $\rho = 1$ and about
$\omega = \Delta_0(1\pm s)$ come together
as the vortex core is approached and they eventually cross each other
before diverging in the opposite direction.
The contours for the critical frequencies, $\omega_c$, are shown in
the bottom frame of Fig.~\ref{f5}. The two ridges that cross are
the solutions of Eq.~(\ref{eq:o5}) for $\phi_c = -\pi/2$ and of
Eq.~(\ref{eq:o6}). The solution of Eq.~(\ref{eq:o5}) for
$\phi_c = \pi/2$ and the plus
sign gives the higher energy ridge shown in the top frame of Fig.~\ref{f5}
(and dash-dotted line in the bottom frame of this figure).

As in the tetragonal case (Fig.~\ref{f4}) the ridge topology
in the LDOS changes considerably as the position of the STM tip
is changed. We illustrate this in the top frame of Fig.~\ref{f6}
where we present results for a winding angle $\beta = 45^\circ$.
Now all four ridges are seen in the entire range
$\nu\le\rho\le 1$ and two cross overs can be observed. The
contours of the critical frequencies $\omega_c$, are shown
in the bottom frame of Fig.~\ref{f6}. A comparison of the top
frame of this figure with the top frame of Fig.~\ref{f4} (tetragonal
case, $\beta = 45^\circ$) makes the splitting of the van Hove
singularity as a result of the subdominant $s$-wave component
of the gap particularly transparent.

\section{Local impurity scattering rate}

\begin{figure}
\caption{Top frame: the impurity scattering rate $\gamma(\rho,\omega)$
in the unitary limit with $\Gamma^+ = 0.1\,$meV as a function of
the distance from the vortex core $\rho$, and frequency $\omega$.
Here the magnetic energy $\nu = 0.1$, ${\bf H}\parallel c$-axis,
and the vortex winding angle $\beta = 0^\circ$, i.e. the STM tip
is moved along the antinodal direction. A system of tetragonal
symmetry is considered. Bottom frame: the same as the top frame
but for a system of orthorhombic symmetry having an effective
mass anisotropy $\alpha = 0.4$ and an $s$-wave component
$s=-0.25$ to model YBCO.}
\label{f7}
\end{figure}
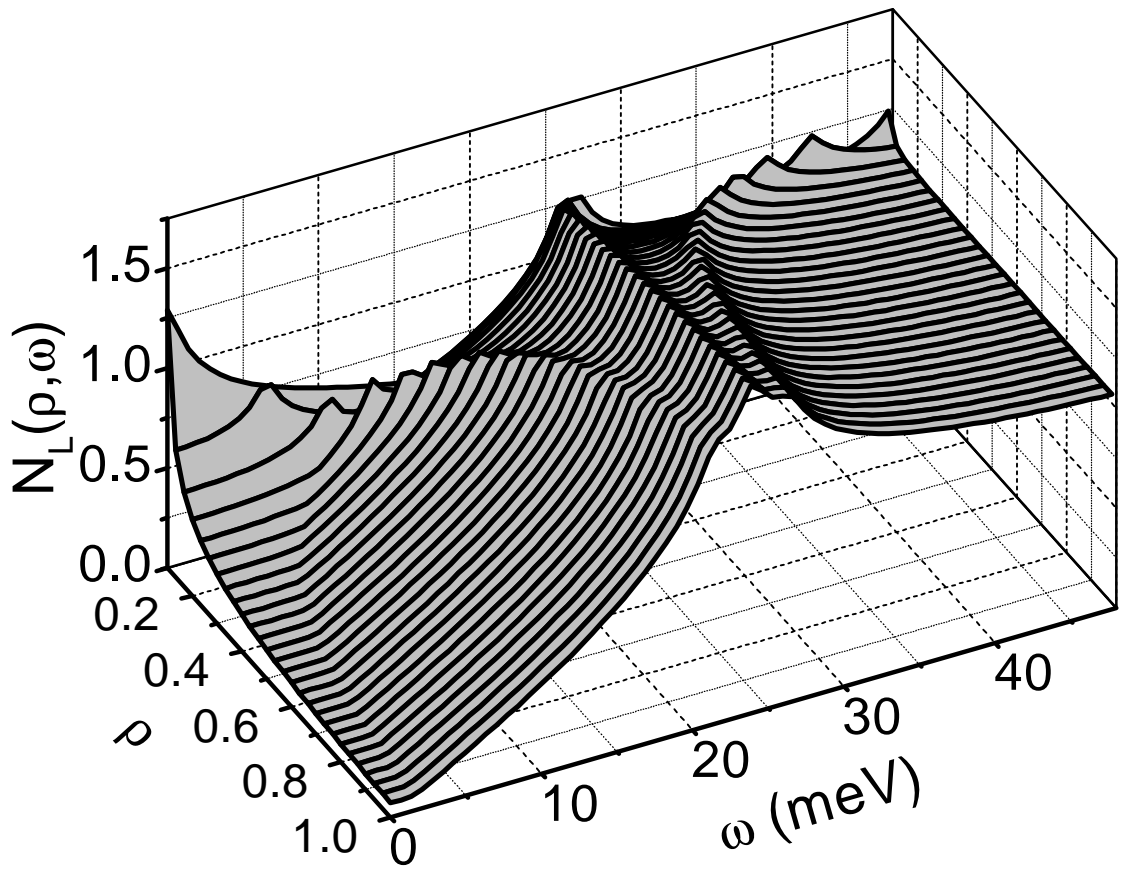
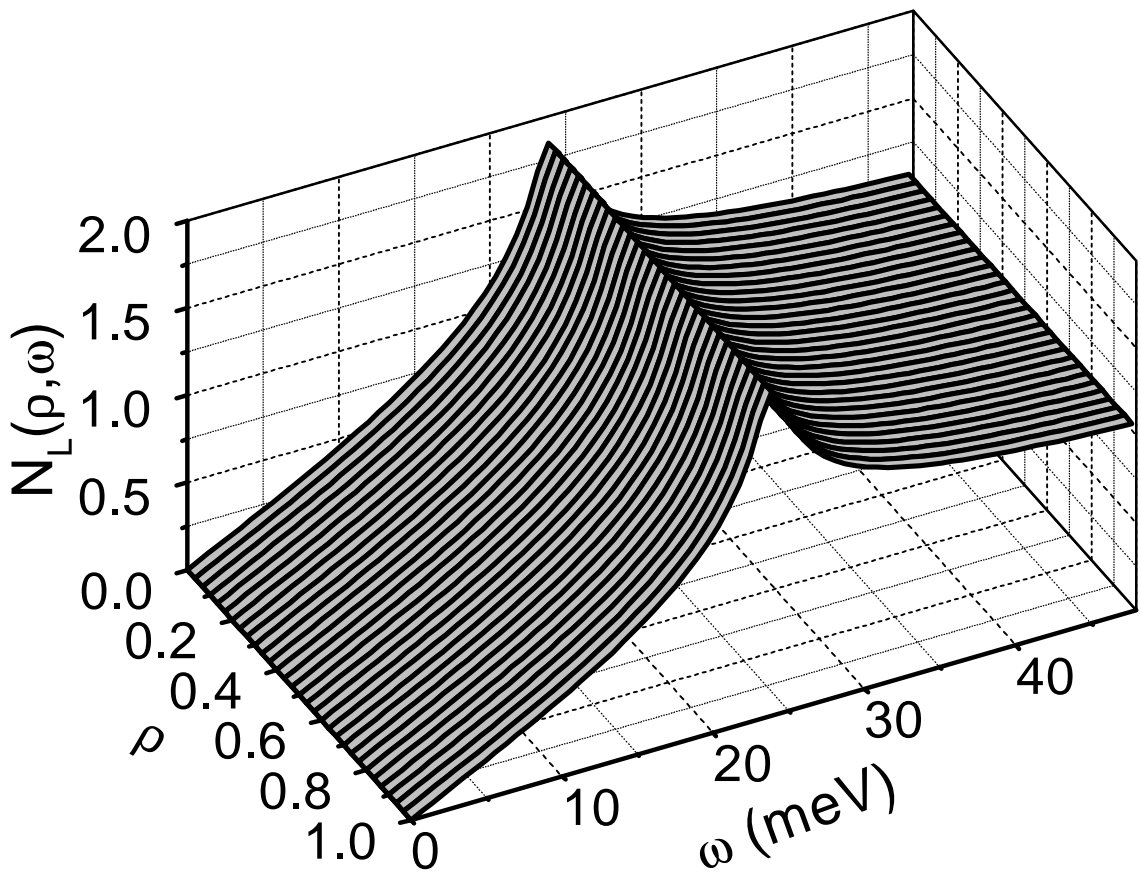
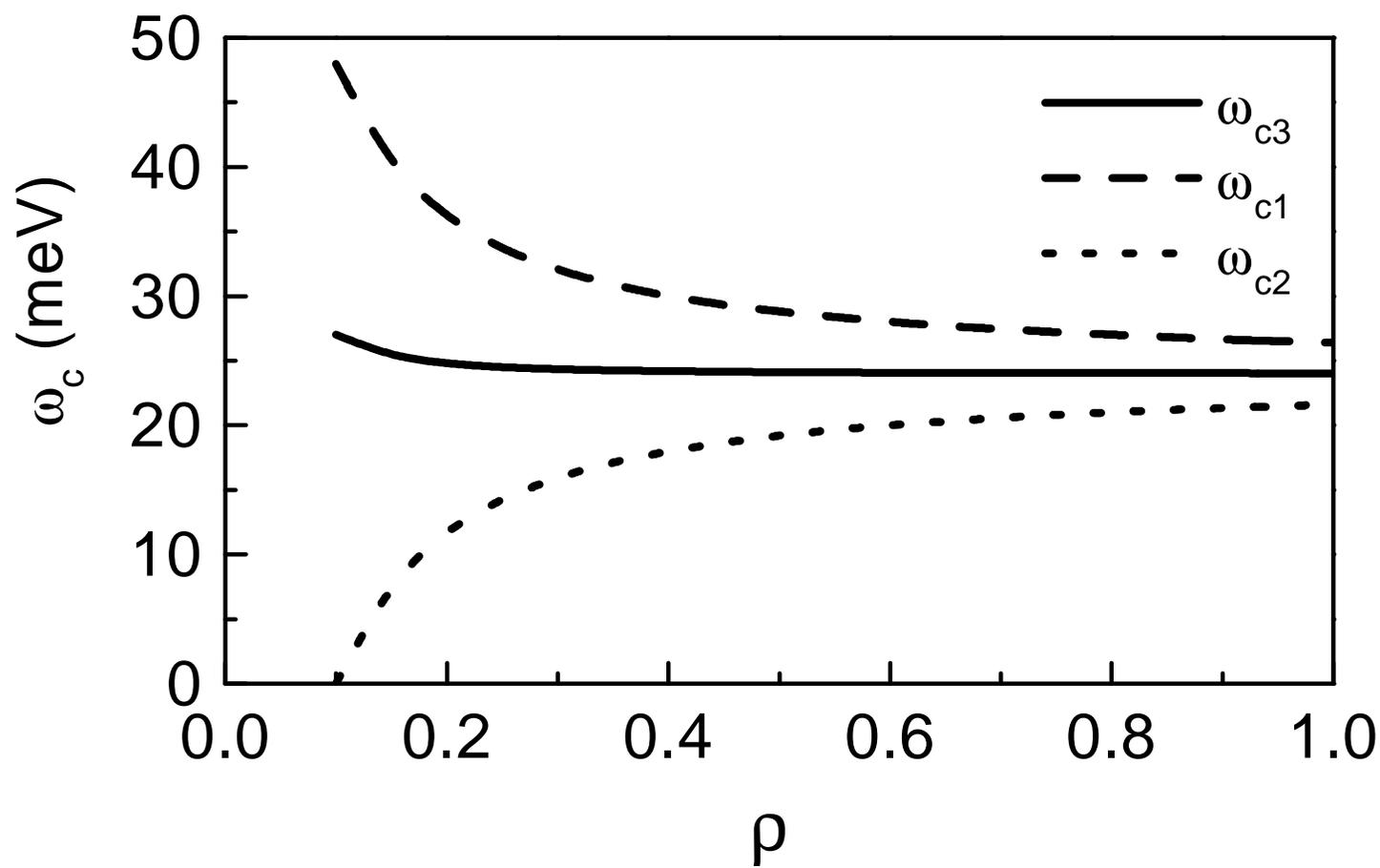
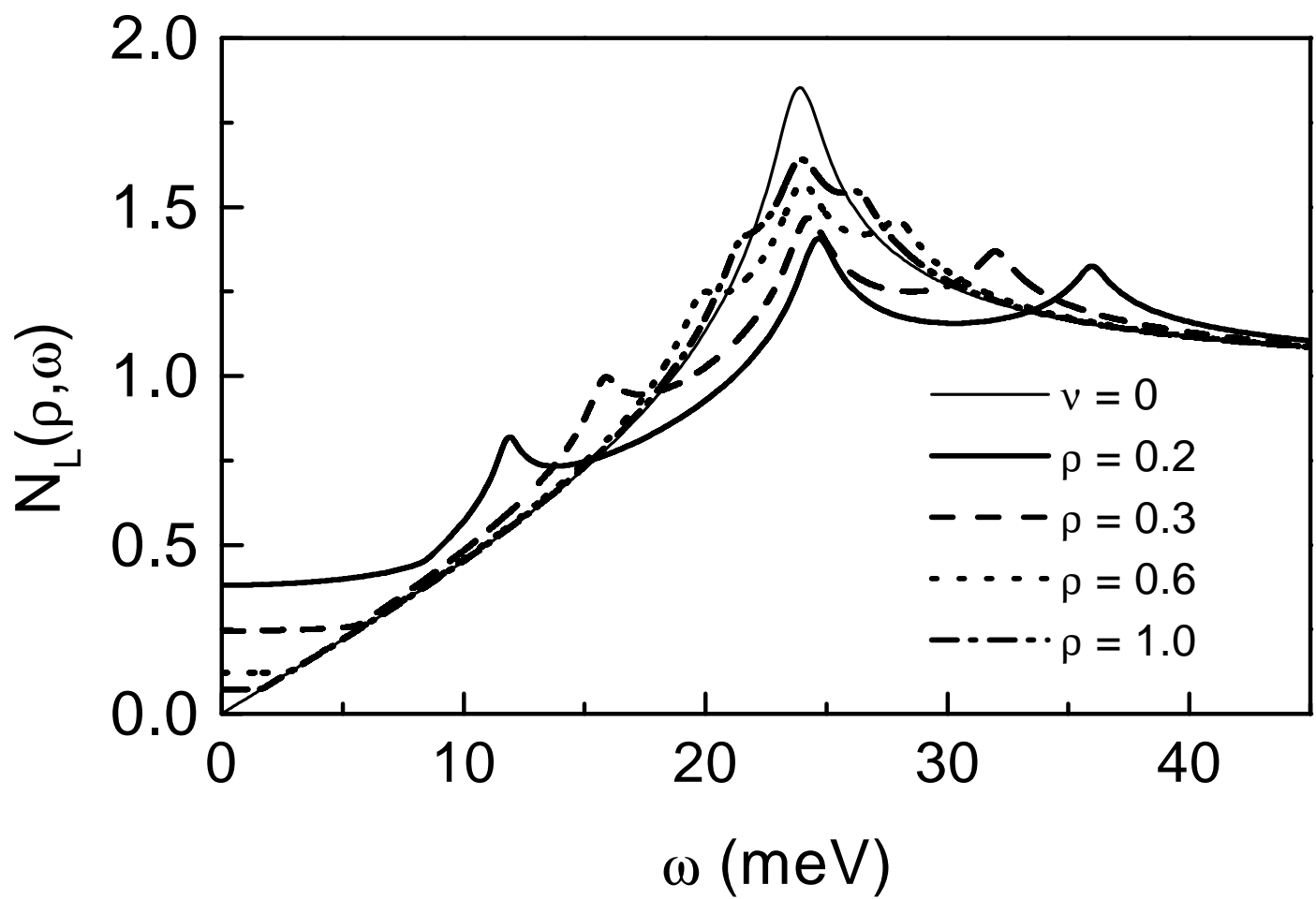
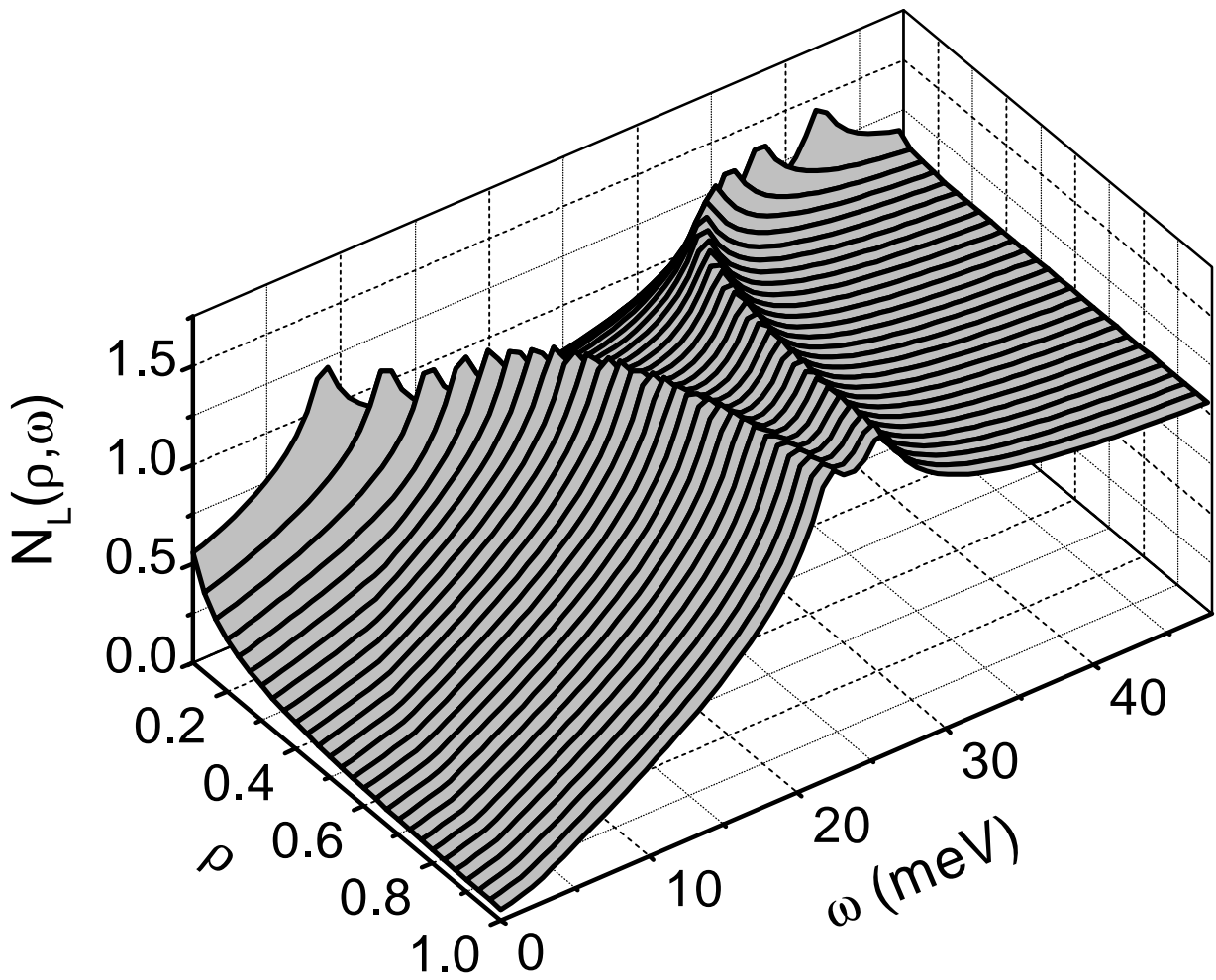
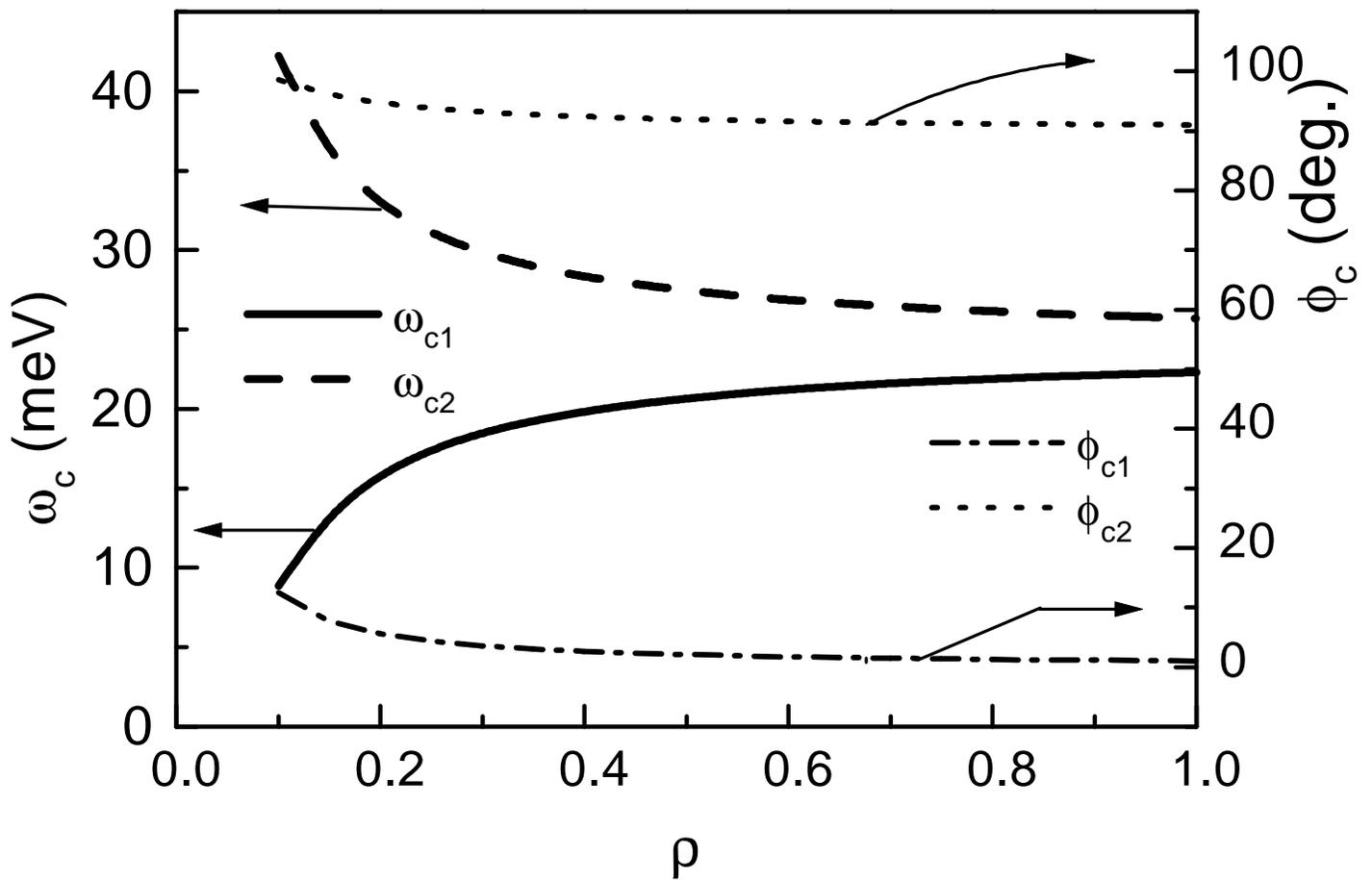
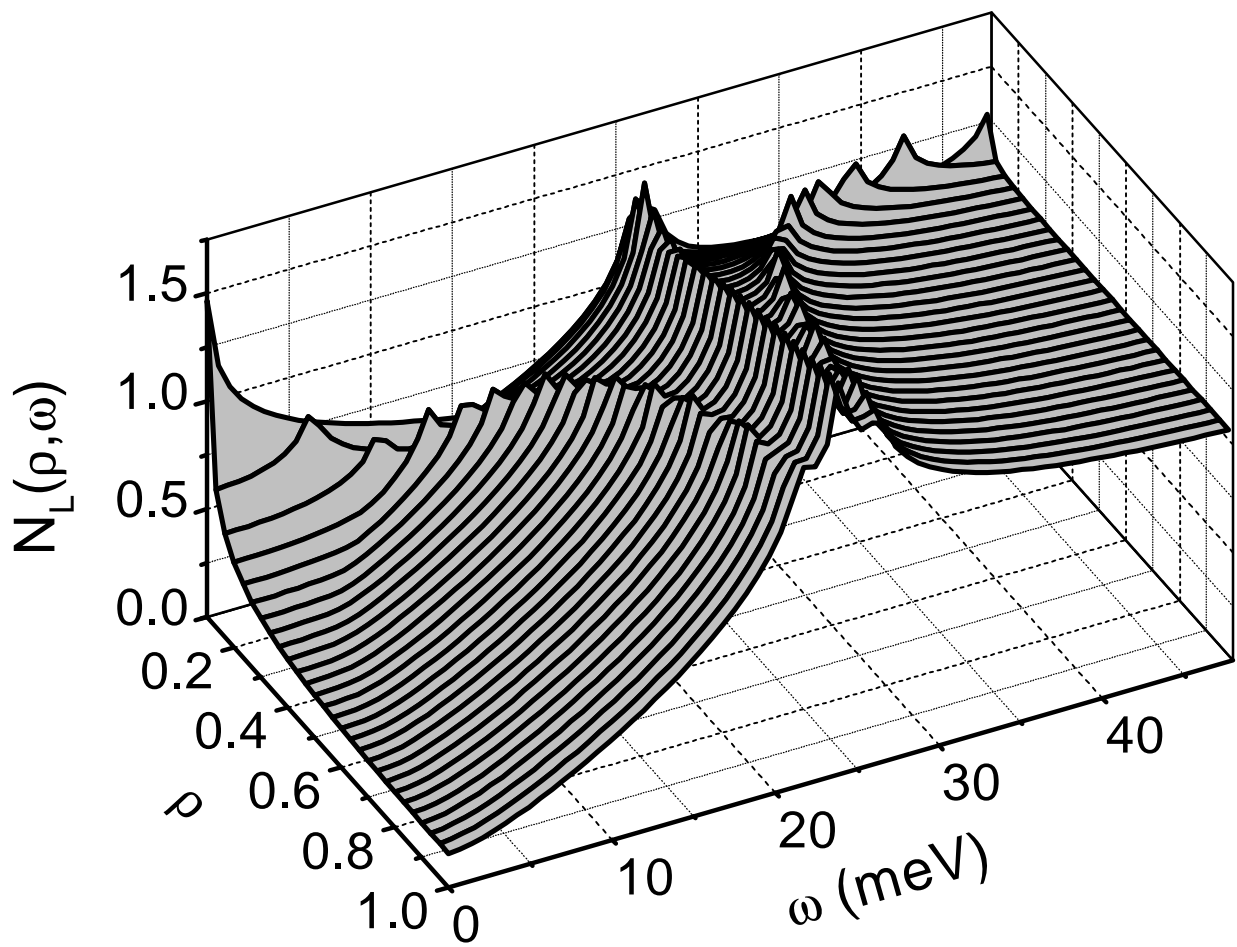
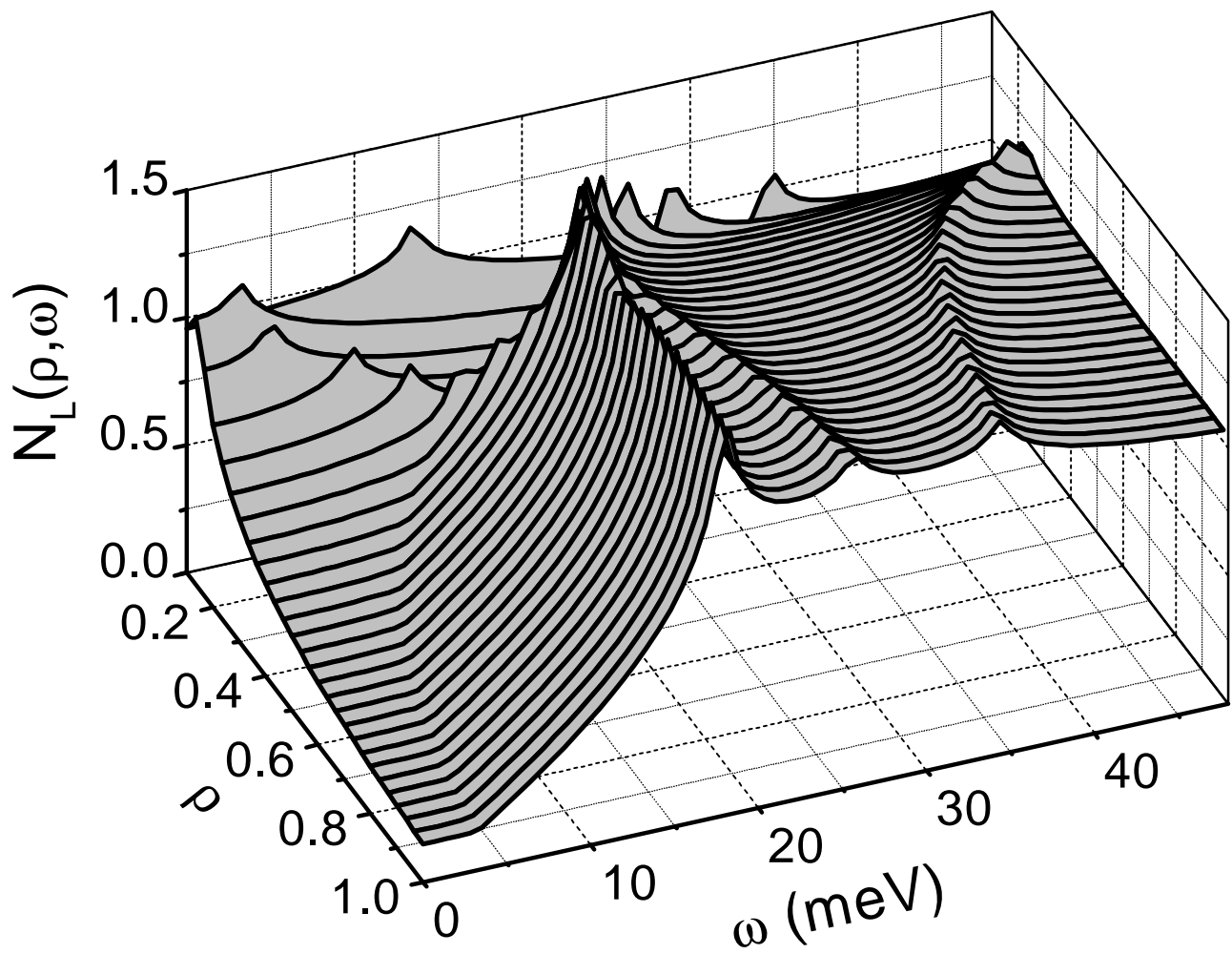
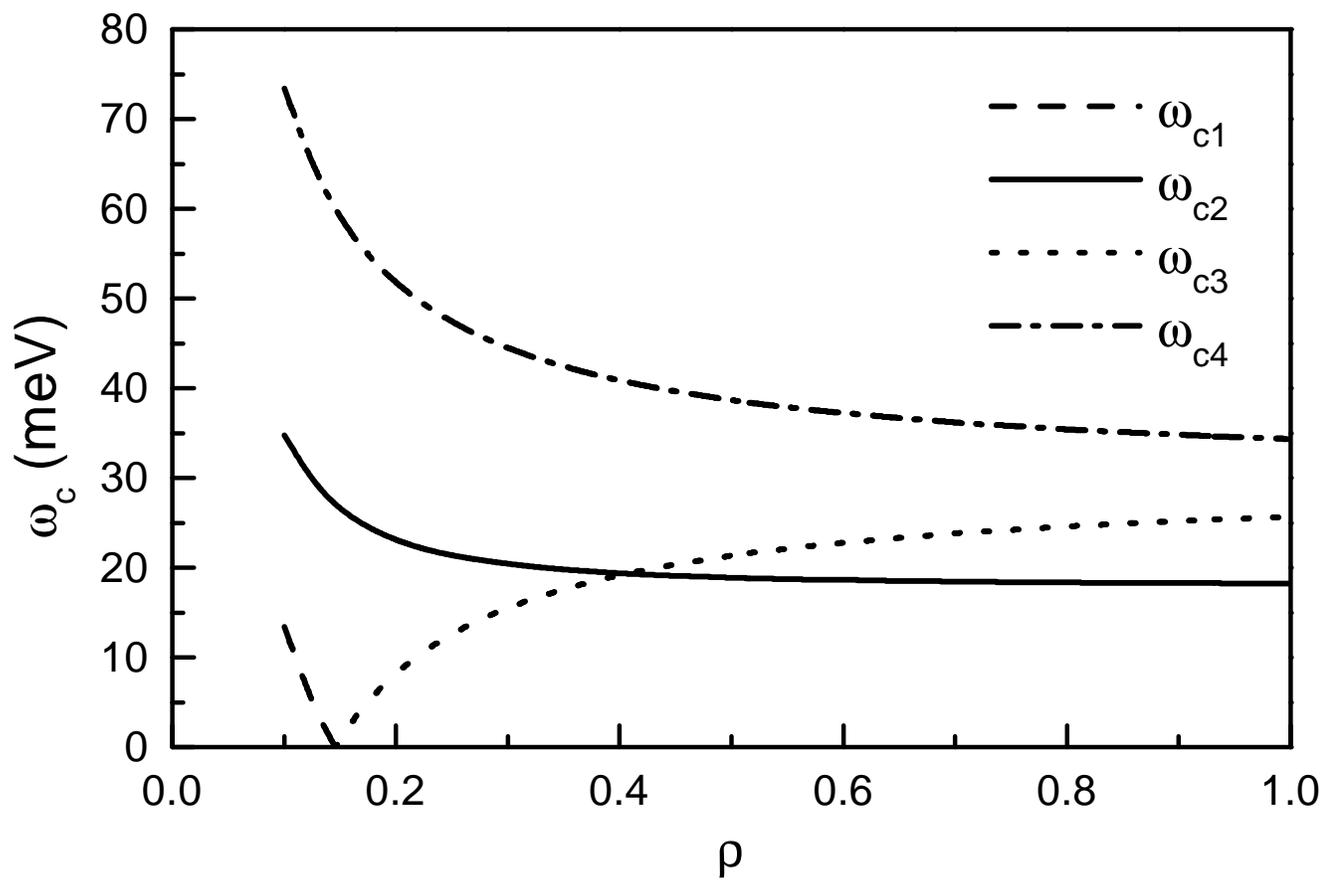
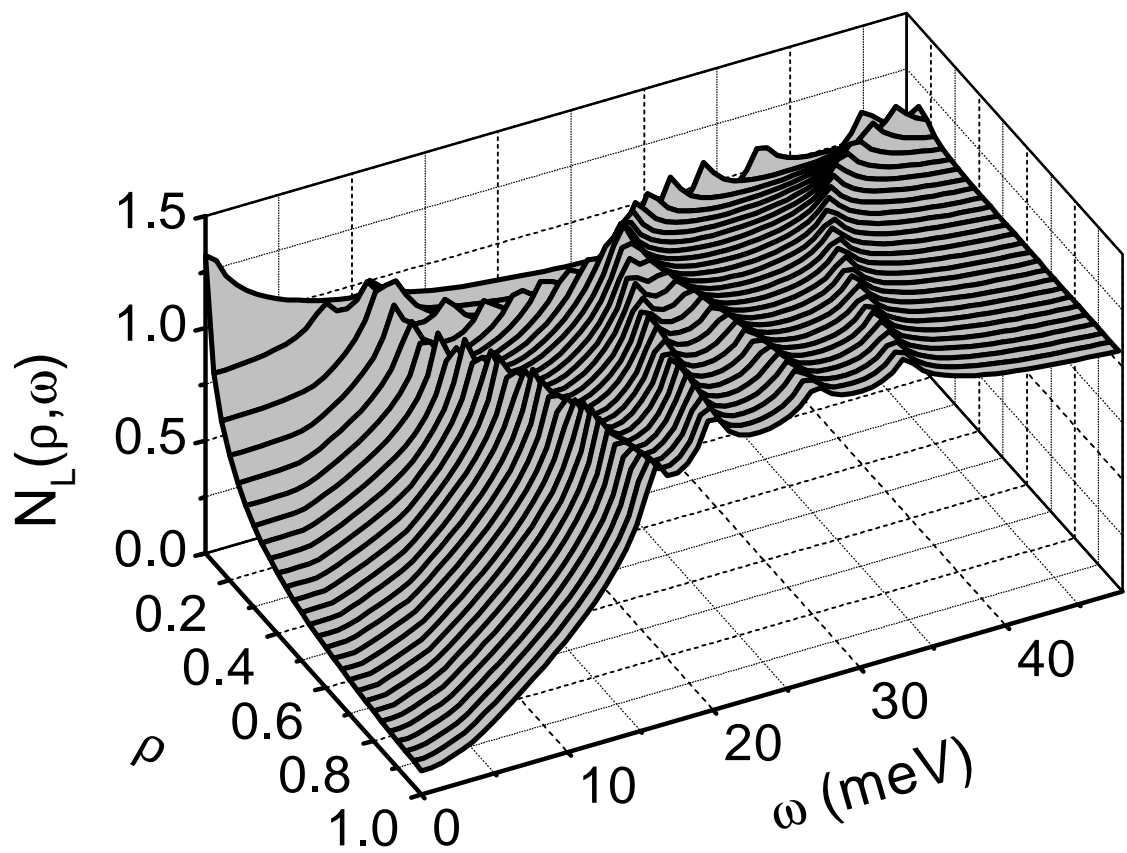
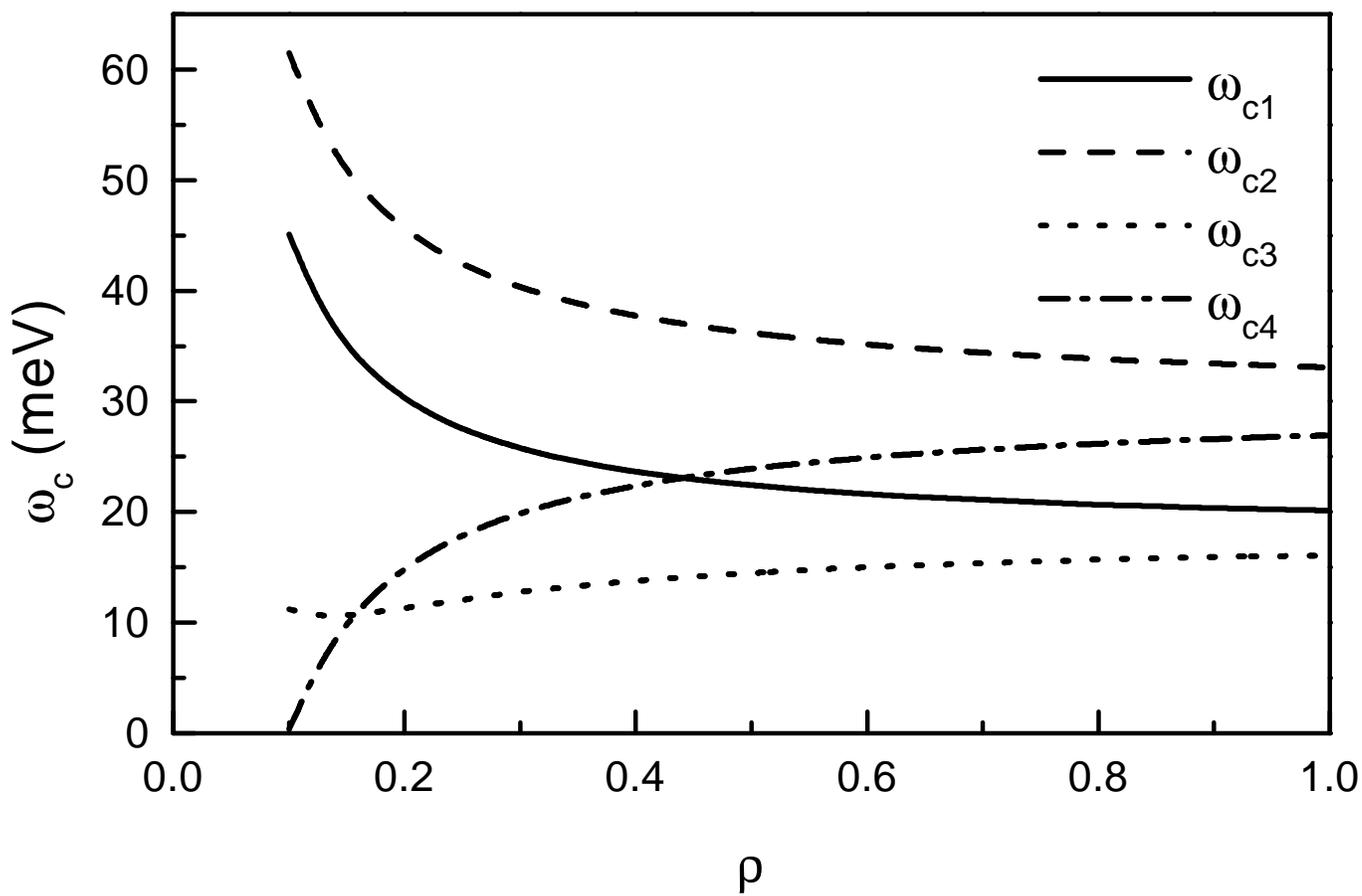
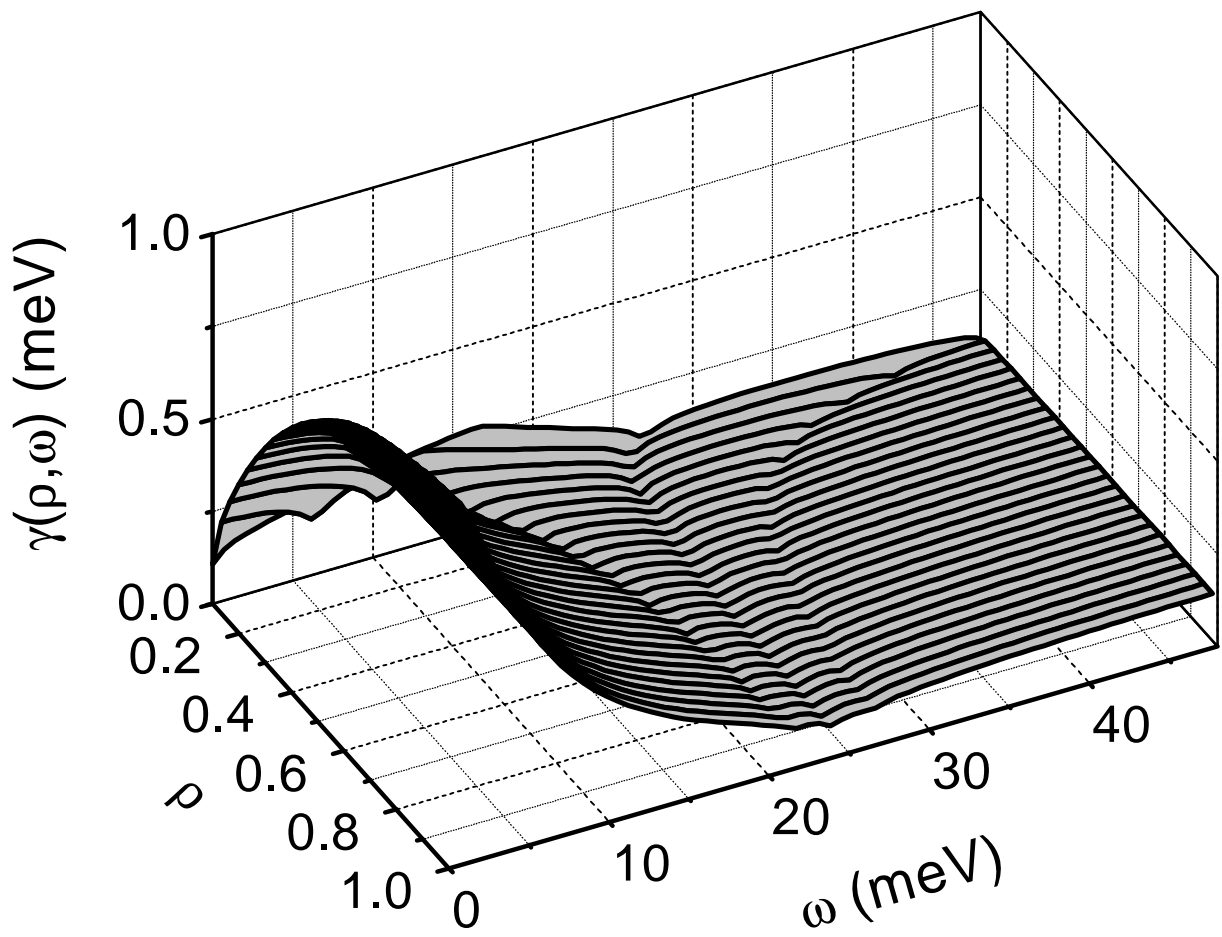
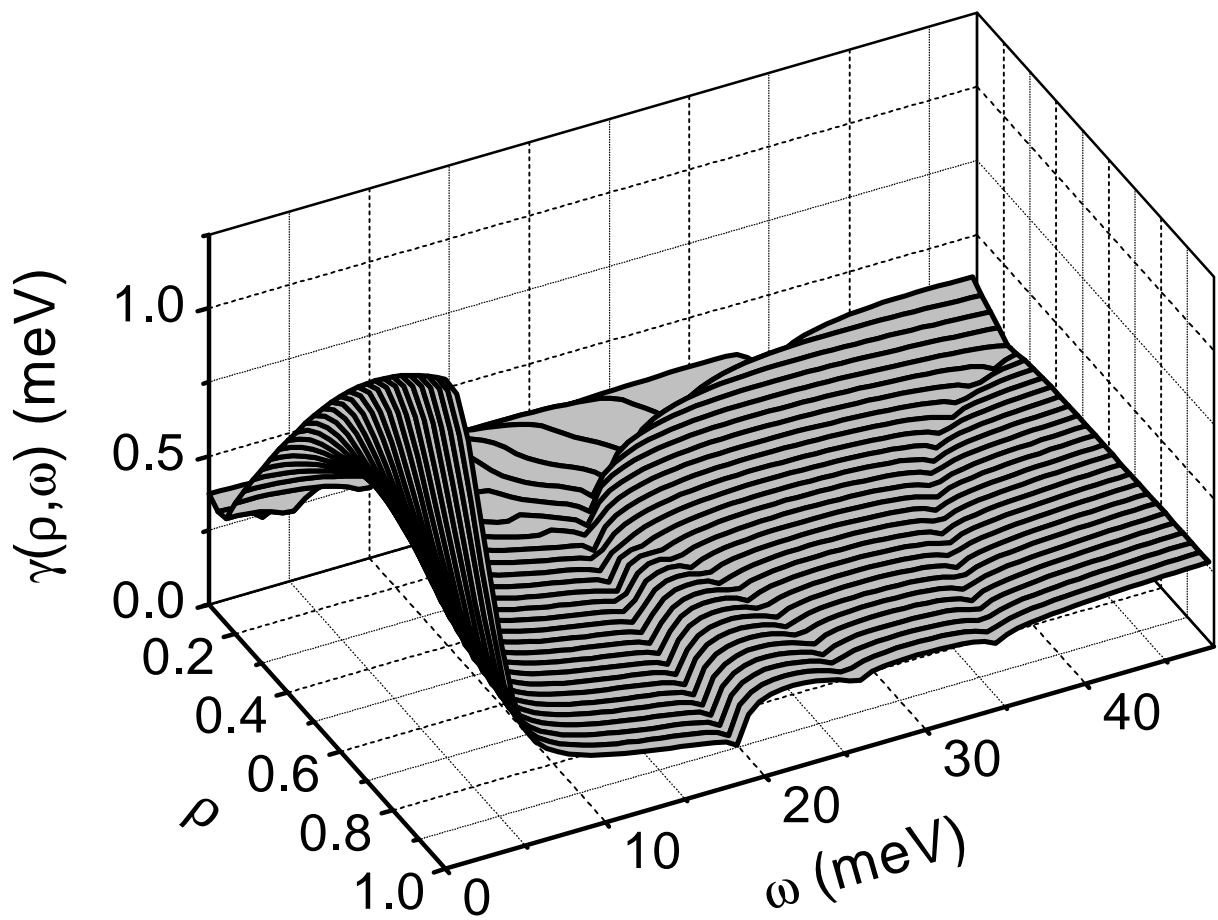

Finally, it is of interest to discuss briefly the local impurity
scattering rates which are modified by the presence of supercurrents.
Similar work has already appeared in a paper by Barash and Svidzinskii.%
\cite{Barash}
In Born approximation $\Im{\rm m}\,\tilde{\omega}(\rho,\beta,\omega)$
given by Eq.~(\ref{eq:4a}) is given by
\begin{equation}
  \Im{\rm m}\,\tilde{\omega}(\rho,\beta,\omega) = 
  \pi t^+\Re{\rm e}\,\Omega(\rho,\beta,\omega),
  \label{eq:8}
\end{equation}
where $\Re{\rm e}\,\Omega(\rho,\beta,\omega)$ is just the LDOS
of Eq.~(\ref{eq:3}) modified by the impurity scattering and
plotted already in the top frame of Fig.~\ref{f1} for the
antinodal direction and in Fig.~\ref{f3} for the nodal direction.
It is clear then that the Doppler shift accounting for the supercurrents
significantly affects the local scattering rates in the same way
as they modify the LDOS. The changes due to the supercurrents
are very different when unitary scattering is considered. In this
case, we find from Eq.~(\ref{eq:4b}),
\begin{equation}
  \Im{\rm m}\,\tilde{\omega}(\rho,\beta,\omega) = 
  \pi\Gamma^+\Re{\rm e}\left\{{\Omega(\rho,\beta,\omega)\over
   \Omega^2(\rho,\beta,\omega)+D^2(\rho,\beta,\omega)}\right\}.
  \label{eq:9}
\end{equation}
For a fixed winding angle $\beta$ defining $\gamma(\rho,\omega) =
\Im{\rm m}\,\tilde{\omega}(\rho,\beta,\omega)$ we have for
$\nu = 0.1$, ${\bf H}\parallel c$-axis, and $\beta = 0^\circ$ the
results shown in the top frame of
Fig.~\ref{f7} where we give a three dimensional
plot of $\gamma(\rho,\omega)$ as a function of distance
$\rho$ from the vortex core and of energy $\omega$ for a
system of tetragonal symmetry. This quantity is strikingly different
from its Born limit counterpart which just corresponds to the
$N_L(\rho,\omega)$ as plotted in the top frame of Fig.~\ref{f1} scaled
by a constant factor of $\pi t^+$.

The $H=0$ case in the unitary limit is quite similar to the
$H\ne 0$ case at $\rho = 1$. The only difference is that for
$H=0$ there is only one shallow dip at $\omega = \Delta_0 =
24\,$meV which does not change position as one moves from
$\rho = 1$ to $\rho = \nu$. It is quite important to notice
that in the limit $\omega\to 0$ $\gamma(\rho,\omega)$
decreases by almost one order of magnitude as one approaches
the vortex core. This means that the supercurrents surrounding
the vortex core will also influence substantially the value
of the universal limit for the transport. 

The equivalent results for a system with orthorhombic
symmetry with an effective mass anisotropy parameter
$\alpha = 0.4$ and an $s$-wave component $s=-0.25$ to the
gap are shown in the bottom frame of Fig.~\ref{f7} to
contrast the results for the system with tetragonal symmetry.
The variation of $\gamma(\rho,\omega)$ in the limit $\omega\to 0$
is less dramatic in this particular case.

It is obvious from these results that
the supercurrents modify significantly the local effective
impurity scattering which depends strongly on $\rho$. This has
already been emphasized by Barash and Svidzinskii\cite{Barash}
in another context.

\section{Conclusion}

In this work we have described the effect of supercurrents around the
vortex core on the LDOS and on the local impurity scattering. We
have included in the work only the effect of the Doppler shift in a
semiclassical approximation on the extended quasiparticles. The
work was motivated by the considerable success such an approach has had
recently in describing the effect of vortex cores on the
thermodynamics and some transport properties of $d$-wave superconductors
for magnetic fields in the range $H_{c1} < H \ll H_{c2}$, between the
lower and upper critical fields. The calculations predict definite
modifications of the expected van Hove singularity at the gap
amplitude $\Delta_0$ in a $d$-wave superconductor. There are
additional ridges which depend on the geometry of the STM
arrangement, i.e. its distance from the vortex core and direction
of the STM with respect to nodal or antinodal direction. They also
depend on orthorhombicity and are modified by impurity scattering.
So far we are not aware of experiments which confirm these predictions.
We hope our work will stimulate more experiments. Of course more
sophisticated approaches to the structure of a vortex core in
a $d$-wave superconductor are possible.\cite{Scho,Ichi1,Ichi2}
We have already mentioned the work of Franz and Te\v{s}anovi\'c.%
\cite{Franz1} They compare results for the LDOS obtained from a
semiclassical approach with results from solutions for a single
vortex within a BdG self consistent approach. For the angular
averages over the vortex winding angle they consider, they do
find good agreement between the two sets of results. An important
additional aspect of these author's work is that they investigate
the possibility that the tunneling matrix element may have angular
dependence and that consequently STM does not directly measure
the LDOS. In the particular model considered, the perpendicular
tunneling is modulated by a $\cos^22\phi$ in-plane dependence
which strongly affects the low energy part of the STM spectrum
because there can be no tunneling right on the diagonal of the
Brillouin zone. This modulation, however, will have less effect
away from the $\omega\sim 0$ region and is not important for
the ridges.
Here our
own emphasis has been on the topology of the secondary van Hove
ridges introduced in the LDOS by supercurrents.
We have derived simple formulas for the position of these ridges
in configuration space, for their dependence on $H$ which varies
with position about the vortex, and have examined the effect of
impurities both in Born and in unitary scattering limit on the
topology of the ridges. We have also considered the modifications
that are introduced when orthorhombicity is introduced within
an anisotropic effective mass model.

\section*{Acknowledgment}

Research supported in part by NSERC (Natural Sciences and
Engineering Research Council of Canada) and CIAR (Canadian
Institute for Advanced Research). We thank E.J.~Nicol for
pointing out the relevance of Ref.~26 to our work.

\newpage
\begin{figure}
\caption{The top frame is the local density of states $N_L(\rho,\omega)$
as a function of energy $\omega$ and the distance from the vortex core
$\rho$ for a magnetic energy $E_H = \nu\Delta_0$ with $\nu = 0.1$ and
a vortex winding angle $\beta = 0^\circ$ along the antinode in the
$d$-wave gap. A small amount of elastic impurity scattering in
Born approximation has been included with $t^+ = 0.1\,$meV. The main
ridge around $24\,$meV is the usual van Hove singularity at the gap
amplitude $\Delta_0$ modified only slightly by the magnetic field $H$.
The other two ridges are due to the Doppler shift of the
semiclassical approximation resultant from the existence of the
supercurrents about the vortex core. The bottom frame is the density
of states when $H = 0$ and is for comparison.}
\label{f1}
\end{figure}
\begin{figure}
\caption{ The top frame
gives the critical values of frequencies $\omega_{ci}, i = 1,2,3$
defining the three ridges in the top frame of Fig.~\ref{f1}.
 The bottom frame represents the same
data as in the top frame of Fig.~\ref{f1} but gives $N_L(\rho,\omega)$ vs. $\omega$ 
for several values of $\rho$, namely $\rho = 0.2$ (solid line),
$\rho = 0.3$ (dashed line), $\rho = 0.6$ (dotted line) and $\rho = 1$
(dash-dotted line). Here, $\rho = 0.1$ corresponds to the vortex core
radius equal to the coherence length $\xi_0$ and $\rho = 1.0$
the vortex cell of Radius $R$. Included is also, for comparison,
$N(\omega)$ for $\nu=0$, i.e. no magnetic field, as a thin solid line.}
\label{f2}
\end{figure}
\begin{figure}
\caption{ The top
frame shows a three dimensional plot of the LDOS $N_L(\rho,\omega)$
as a function of distance from the vortex core $\rho$ in the
vortex unit cell, and of frequency $\omega$. Here
the vortex winding angle
is $\beta = 45^\circ$, i.e. the STM tip is moved along the nodal
direction, and the magnetic energy is $\nu = 0.1$. The bottom
frame gives the critical frequency $\omega_{ci}, i = 1,2$ and
the corresponding critical angles $\phi_{ci}$
that locate the ridges in the top frame as a function of the
distance from the vortex core $\rho$, for $\nu = 0.1$,
${\bf H}\parallel c$-axis, $\beta = 45^\circ$.}
\label{f3}
\end{figure}
\begin{figure}
\caption{A three dimensional plot of the LDOS as a function of
the distance $\rho$ from the vortex core in the vortex unit cell,
and of frequency $\omega$. Here the magnetic energy is $\nu = 0.1$,
${\bf H}\parallel c$-axis, and the vortex winding angle
$\beta = 0^\circ$ (antinodal direction). Unitary impurity scattering
is included with $\Gamma^+ = 0.1\,$meV as compared to the
top frame of Fig.~\ref{f1} where the Born limit is used instead.}
\label{f4}
\end{figure}
\begin{figure}
\caption{Top frame: a three dimensional plot of the LDOS $N_L(\rho,\omega)$
as a function of distance from the vortex core $\rho$ in the vortex unit
cell, and of frequency $\omega$. Here the vortex winding angle
$\beta = 0$ (antinodal direction) and ${\bf H}\parallel c$-axis, i.e.
perpendicular to the copper oxide planes. The mass anisotropy
$\alpha = 0.4$ and an $s$-wave component $s = -0.25$ are included
to model the orthorhombicity of YBCO. Born impurity scattering with 
$t^+=0.1\,$meV is also included. The bottom frame gives the critical
frequencies $\omega_c$ defining the van Hove ridges in the top frame.}
\label{f5}
\end{figure}
\begin{figure}
\caption{The top frame shows a three dimensional plot of the LDOS
$N_L(\rho,\omega)$ as a function of distance from the vortex core
$\rho$ in the vortex unit cell, and of frequency $\omega$. Here
the vortex winding angle $\beta = 45^\circ$
and ${\bf H}\parallel c$-axis. The effective mass anisotropy
$\alpha = 0.4$ and an $s$-wave component $s=-0.25$ are included
to model the orthorhombicity of YBCO. Born impurity scattering with
$t^+ = 0.1\,$meV is also included. The bottom frame gives the critical
frequencies $\omega_c$ defining the van Hove singularities in
the top frame.}
\label{f6}
\end{figure}
\end{document}